\documentclass[aps,prd,twocolumn,amsmath,showpacs,amssymb,superscriptaddress,nofootinbib]{revtex4}
\usepackage{graphicx}
\usepackage{dcolumn}
\usepackage{bm}
\usepackage{amssymb,amsmath}
\usepackage{latexsym}
\usepackage{color}
\usepackage[normalem]{ulem} 
\usepackage{cancel}
\allowdisplaybreaks
\newcommand{\Caltech}{\affiliation{Theoretical Astrophysics 350-17,
    California Institute of Technology, Pasadena, California 91125, USA}}
\newcommand{\be}{\begin{equation}}
\newcommand{\ee}{\end{equation}}
\newcommand{\ba}{\begin{eqnarray}}
\newcommand{\ea}{\end{eqnarray}}
\newcommand{\bma}{\begin{pmatrix}}
\newcommand{\ema}{\end{pmatrix}}

\begin{document}

\title{Classifying the Isolated Zeros of Asymptotic Gravitational Radiation by Tendex and Vortex Lines}

\author{Aaron Zimmerman} \Caltech
\author{David A.\ Nichols} \Caltech
\author{Fan Zhang} \Caltech
\date{\today}

\begin{abstract}
A new method to visualize the curvature of spacetime was recently proposed. 
This method finds the eigenvectors of the electric and magnetic components of 
the Weyl tensor and, in analogy to the field lines of electromagnetism, uses the eigenvectors' integral curves to illustrate the spacetime curvature. 
Here we use this approach, along with well-known topological properties of 
fields on closed surfaces, to show that an arbitrary, radiating, 
asymptotically flat spacetime must have points near null infinity where the 
gravitational radiation vanishes. 
At the zeros of the gravitational radiation, the field of integral curves
develops singular features analogous to the critical points of a vector field.
We can, therefore, apply the topological classification of singular points of unoriented lines as a method to describe the radiation field.
We provide examples of the structure of these points using linearized gravity
and discuss an application to the extreme-kick black-hole-binary merger.
\end{abstract}
\maketitle
 
\section{Introduction}
\label{sec:Intro}

A recent study \cite{Owen2011} proposed a method for visualizing 
spacetime curvature that is well-suited for studying spacetimes evolved from initial data using numerical-relativity codes. 
The method first projects the Riemann 
curvature tensor $R_{\mu \nu \sigma \rho}$ into a spatial slice, thereby 
splitting it into two symmetric, trace-free spatial tensors, $\bm{\mathcal E}$ 
and $\bm{\mathcal B}$ (see e.g. \cite{Maartens1998} and the references therein).
These tensors are the spacetime-curvature analogs of the electric and magnetic fields in 
Maxwell's theory. 
The electric tensor $\bm {\mathcal E}$ is familiar; it is the tidal field 
in the Newtonian limit. 
The frame-drag field $\bm{\mathcal B}$ (the magnetic curvature tensor) 
describes the differential frame dragging of spacetime. 
The eigenvectors of the tidal field provide the preferred directions 
of strain at a point in spacetime, and its eigenvalues give the magnitude of 
the strain along those axes. Similarly, the eigenvectors of the frame-drag 
field give preferred directions of 
differential precession of gyroscopes, and their eigenvalues give 
the magnitude of this precession \cite{Owen2011, Estabrook1964, PRD1}. 

The study \cite{Owen2011} then proposed using the integral curves of these 
eigenvectors as a way to visualize the curvature of spacetime.
Three orthogonal curves associated with $\bm{\mathcal E}$, called tendex 
lines, pass through each point in spacetime. 
Along each tendex line there is a corresponding eigenvalue, which is called 
the tendicity of the line.
For the tensor $\bm{\mathcal B}$, there is a second set of three orthogonal 
curves, the vortex lines, and their corresponding eigenvalues, the vorticities.
These six curves are analogous to the field lines of electromagnetism, and the 
six eigenvalues to the electric and magnetic field strengths.
The tendex and vortex lines, with their corresponding vorticities and 
tendicities, represent very different physical phenomena from field lines
of electromagnetism; they allow one to visualize the aspects of spacetime
curvature associated with tidal stretching and differential frame-dragging.
In addition, each set of curves satisfies the constraint that its eigenvalues 
sum to zero at every point, since $\bm{\mathcal E}$ and $\bm{\mathcal B}$ are 
trace-free.

Wherever the eigenvector fields are well-behaved, the tendex and vortex lines 
form extended, continuous fields of lines in a spatial slice. 
At points where two (or more) eigenvectors have the same eigenvalue, the 
eigenvectors are said to be degenerate. 
Any linear combination of the degenerate eigenvectors at these points is still 
an eigenvector with the same eigenvalue; therefore, the span of these
eigenvectors forms a degenerate subspace.
Singular features can appear at points of degeneracy, where many lines 
intersect, terminate, or turn discontinuously.
The topology of unoriented fields of lines and their singular points has 
been studied both in the context of general relativity and elsewhere.
For example, Delmarcelle and Hesselink \cite{DelmarcelleHesselink1994} studied
the theory of these systems and applied them to real, symmetric two-dimensional
tensors. In the context of relativity, Penrose and Rindler \cite{PenroseRindler2}
examined the topology of unoriented lines, or ridge systems, to characterize 
the principle null directions about single points in spacetime.
Finally, Penrose \cite{Penrose1979} also applied the study of ridge systems to human handprint and fingerprint patterns.

In this paper, we focus on the vortex and tendex lines and their singular
points far from an isolated, radiating source.
In Sec. \ref{sec:GravWaves}, we show that two of the vortex and tendex 
lines lie on a sphere (the third, therefore, is normal to the sphere), and 
that the vortex and tendex lines have the same eigenvalues. 
Moreover, the two eigenvalues on the sphere have opposite sign, and the
eigenvalue normal to the sphere has zero eigenvalue.
This implies that the only singular points in the lines occur when all
eigenvalues vanish (i.e.\ when the curvature is exactly zero at the point, and all three eigenvectors are degenerate).

In Sec. \ref{sec:Topology} we employ a version of the Poincar\'e-Hopf
theorem for fields of integral curves to argue that there must be singular
points where the curvature vanishes. 
Penrose, in a 1965 paper \cite{Penrose1965}, made a similar observation.
There, he notes in passing that gravitational radiation must vanish for 
topological reasons, although he does not discuss the point any further. 
Here we show that the topological classification of singular points of ridge 
systems can be applied to the tendex and vortex lines of gravitational 
radiation.
This allows us to make a topological classification of the zeros of
the radiation field.

In Sec. \ref{sec:Examples}, we visualize the tendex and vortex lines
of radiating systems in linearized gravity. 
We begin with radiation from a rotating mass-quadrupole moment, the dominant
mode in most astrophysical gravitational radiation.
We then move to an idealized model of the ``extreme-kick'' configuration 
(an equal-mass binary-black-hole merger with spins antialigned in the orbital 
plane \cite{CampanelliLousto2007}).
As we vary the magnitude of the spins in the extreme-kick configuration, we 
can relate the positions of the singular points of the tendex and vortex 
patterns to the degree of beaming of gravitational waves. 
We also visualize the radiation fields of individual higher-order multipole 
moments, which serve, primarily, as examples of patterns with a large number 
of singularities.
Astrophysically, these higher multipoles would always be accompanied by 
a dominant quadrupole moment; we also, therefore, look at a superposition of
multipoles.
Since the tendex lines depend nonlinearly upon the multipoles, it is not
apparent, {\it a priori}, how greatly small higher multipoles will change the
leading-order quadrupole pattern.
Nevertheless, we see that for an equal-mass black-hole binary, higher
multipoles make only small changes to the tendex line patterns.
Finally, we discuss our results in Sec.\ \ref{sec:Conclusions}.

Throughout this paper we use Greek letters for spacetime coordinates in a
coordinate basis and Latin letters from the beginning of the alphabet for
spatial indices in an orthonormal basis.
We use a spacetime signature $(-+++)$ and a corresponding normalization
condition for our tetrad.
We will use geometric units, in which $G=c=1$.

We will also specialize to vacuum spacetimes, where the
Riemann tensor is equal to the Weyl tensor $C_{\mu \nu \rho \sigma}$.
To specify our slicing and to compute $\bm{\mathcal E}$ and $\bm{\mathcal B}$,
we use a hypersurface-orthogonal, timelike unit vector, $\bm {e}_0$,
which we choose to be part of an orthonormal tetrad, 
$(\bm{e}_0, \bm{e}_1, \bm{e}_2, \bm{e}_3)$. 
We then perform a $3+1$ split of the Weyl tensor by projecting it and its 
Hodge dual ${}^*C_{\mu \nu \rho \sigma} =\frac 12 \epsilon_{\mu \nu}{}^{\alpha \beta} C_{\alpha \beta \sigma \rho}$ into this basis,
\ba
\label{RiemSplitE}
\mathcal E_{ab} &=& C_{a \mu b \nu} e_0{}^{\mu} e_0{}^{\nu} \,, \\
\label{RiemSplitB}
\mathcal B_{ab} &=& - ^*C_{a \mu b \nu} e_0{}^{\mu} e_0{}^{\nu} \, .
\ea
Here our convention for the alternating tensor is that $\epsilon_{0123} = +1$ in an orthonormal basis. Note that, while the sign convention on $\bm {\mathcal B}$ is not standard (see e.g. \cite{Stephani}), it has the advantage that $\bm {\mathcal E}$ and $\bm {\mathcal B}$ obey constraints and evolution equations under the $3+1$ split of spacetime that are directly analogous to Maxwell's equations in electromagnetism \cite{Maartens1998,PRD1}. After the projection, we will solve the eigenvalue problem for the tensors $\bm{\mathcal E}$ and $\bm{\mathcal B}$ in the orthonormal basis,
\be
\label{EigenvectorEqn}
\mathcal{E}_{ab} v^{b} = \lambda v_a \, ,
\ee
and we will then find their streamlines in a coordinate basis via the
differential equation relating a curve to its tangent vector,
\begin{equation}
\label{StreamlineEqn}
\frac{dx^\mu}{ds} = v^a e_a{}^\mu \, . 
\end{equation}
Here $s$ is a parameter along the streamlines.

\section{Gravitational Waves Near Null Infinity}
\label{sec:GravWaves}

Consider a vacuum, asymptotically flat spacetime that contains gravitational
radiation from an isolated source. 
We are specifically interested in the transverse modes of radiation on a large
sphere $S$ near future null infinity. 
To describe these gravitational waves, we use an orthonormal tetrad  
$(\bm{e}_0, \bm{e}_1, \bm{e}_2, \bm{e}_3)$, with $\bm{e}_0$ timelike and  
$\bm {e}_2, \bm{e}_3$ tangent to the sphere, and we associate with this tetrad 
a corresponding complex null tetrad, 
\ba
\label{OrthonormalBasis}
\bm{l} = \frac{1}{\sqrt{2}} ( \bm{e}_0 + \bm{e}_1) \,, & \qquad & \bm{n} = \frac{1}{\sqrt{2}} ( \bm{e}_0 - \bm{e}_1) \,,
\nonumber \\
\bm{m} = \frac{1}{\sqrt{2}} ( \bm{e}_2 + i \bm{e}_3) \,, & \qquad & \bm{\bar m} = \frac{1}{\sqrt{2}} ( \bm{e}_2 - i \bm{e}_3) \,.
\ea
Here, $\bm l$ is tangent to outgoing null rays that pass through $S$ and 
strike a sphere at null infinity.
We enforce that the null tetrad is parallelly propagated along these rays, and 
that it is normalized such that $l_\mu n^\mu = - m_\mu \bar m^\mu = -1$ (all other inner products of the null tetrad vanish). 
With these rays, we can associate Bondi-type coordinates (see e.g. \cite{Bondi1962,Tamburino1966}) 
on a sphere at future null infinity with those on $S$. 
The timelike vector $\bm{e}_0$ specifies our spatial slicing in this 
asymptotic region. 
When the orthonormal and null tetrads are chosen as in Eq.\ 
\eqref{OrthonormalBasis}, $\bm{\mathcal E}$ and $\bm{\mathcal B}$ are related 
to the complex Weyl scalars \cite{NewmanPenrose1962}. 
With the Newman-Penrose conventions appropriate to our metric signature (see, e.g., \cite{Stephani}), and our convention in  Eq.\ \eqref{RiemSplitB}, one can show that
\be
\label{QEqn}
\mathcal E_{ab} + i \mathcal B_{ab} =
\begin{pmatrix}
2 \Psi_2 & \Psi_3 - \Psi_1&  i (\Psi_1 + \Psi_3) \\  
* & \frac{\Psi_0+ \Psi_4 }{2}- \Psi_2 & \frac{i(\Psi_4-\Psi_0)}{2} \\
* & * & - \frac{\Psi_0+\Psi_4}{2} - \Psi_2 
\end{pmatrix} \,,
\ee
where $*$ indicates entries that can be inferred from the symmetry of $\bm{\mathcal E}$ and $\bm{\mathcal B}$.

In an asymptotically flat spacetime, the peeling theorem \cite{NewmanPenrose1962} ensures that 
$\Psi_4 \sim r^{-1}$ (with $r$ an affine parameter along the rays), and that 
the remaining Weyl scalars fall off with progressively higher powers of $r$,
$ \Psi_3 \sim r^{-2}, \Psi_2 \sim r^{-3}, \Psi_1 \sim r^{-4}$, and 
$\Psi_0 \sim r^{-5}$ . 
Asymptotically, only $\Psi_4$ contributes to $\bm{\mathcal E}$ and 
$\bm{\mathcal B}$, 
\be
\label{PlaneWaveQ}
\mathcal E_{ab} +  i \mathcal B_{ab}  = \frac{1}{2} 
\begin{pmatrix}
0 &0&  0\\  
0 & \Psi_4 & i\Psi_4 \\
 0 &i \Psi_4 & -\Psi_4 
\end{pmatrix}
\, .
\ee
We see immediately that one eigenvector of both  $\bm{\mathcal E}$ and 
$\bm{\mathcal B}$ is the ``radial'' basis vector $\bm{e}_1$, 
with vanishing eigenvalue. The remaining $2 \times 2$ block is transverse and traceless, and the eigenvectors in this subspace have a simple analytical solution.
The eigenvalues are $\lambda_\pm=\pm |\Psi_4|/2$ for both tensors, and the 
eigenvectors of $\bm{\mathcal E}$ have the explicit form
\ba
\label{Eigenvecs}
{\bm v_\pm} & = & \frac{-\mathcal{E}_{23} \bm{e}_2 +(\mathcal E_{22} - \lambda_\pm)  \bm{e}_3}{\sqrt{\mathcal E_{23}^2+(\mathcal E_{22} - \lambda_\pm)^2 }}
\nonumber \\
 & =&   \frac{{\rm Im} \Psi_4 \bm{e}_2 + ({\rm Re} \Psi_4 \mp |\Psi_4|) \bm{e}_3}{\sqrt{ ({\rm Im} \Psi_4 )^2+ ({\rm Re} \Psi_4 \mp|\Psi_4|)^2}} \,.
\ea

The eigenvectors of $\bm{\mathcal B}$ are locally rotated by $\pm \pi/4$ with 
respect to those of $\bm{\mathcal E}$ \cite{PRD1}.
As a result, although the global geometric pattern of vortex and tendex
lines may differ, their local pattern and their topological properties on $S$ 
will be identical. 
Moreover, when the eigenvalues of $\bm{\mathcal E}$ (the tendicity of the 
corresponding tendex line) vanish, so must those of $\bm{ \mathcal B}$
(the vorticity of the vortex lines). 
In arguing that the radiation must vanish, we can, therefore, focus on the 
tendex lines on $S$ without loss of generality. 
Physically, however, both the vortex and the tendex lines are of interest.
Similarly, since the two sets of tendex lines on $S$ have equal and opposite 
eigenvalue and are orthogonal, we need only consider the properties of a 
single field of unoriented lines on $S$ in order to describe the topological 
properties of all four tendex and vortex lines on the sphere.
Note that thus far we leave the coordinates $(x^2, x^3)$ on $S$ unspecified. 
We will assume that these coordinates are everywhere nonsingular, for instance 
by being constructed from two smooth, overlapping charts on $S$. 

\section{The Topology of Tendex Patterns Near Null Infinity}
\label{sec:Topology}

Before investigating the properties of the tendex lines on $S$, 
we first recall a few related properties of vector fields on a 2-sphere.
A well-known result regarding vector fields on a sphere is the
``hairy-ball theorem.''
This result states, colloquially, that if a sphere is covered with hairs
at each point, the hair cannot be combed down everywhere without 
producing cowlicks or bald spots.
The hairy-ball theorem is a specific illustration of the Poincar\'{e}-Hopf theorem, applied to a $2$-sphere. On a $2$-sphere, this theorem states that the sum of the indices of the zeros of a 
vector field must equal the Euler characteristic, $\chi$, of the sphere, specifically $\chi = 2$. 
The index of a zero of a vector field (also called a singular point) can be 
found intuitively by drawing a small circle around the point and traveling 
once around the circle counterclockwise.
The number of times the local vector field rotates counterclockwise through 
an angle of $2 \pi$ during this transit is the index. 
More precisely, we can form a map from the points near a zero of the vector 
field to the unit circle. 
To do this consider a closed, oriented curve in a neighborhood of the zero, 
and map each point on the curve to the unit circle by associating the 
direction of the vector field at that point with a particular point on the 
unit circle. 
The index is the degree of the map (the number of times the map covers the 
circle in a positive sense). 
For the zero of a vector field, the index is a positive or negative integer, 
because we must return to the starting point on the unit circle as we finish our circuit of the curve around the zero.

\begin{figure}[tb]
\centering
\includegraphics[width = 3.325 in, keepaspectratio]{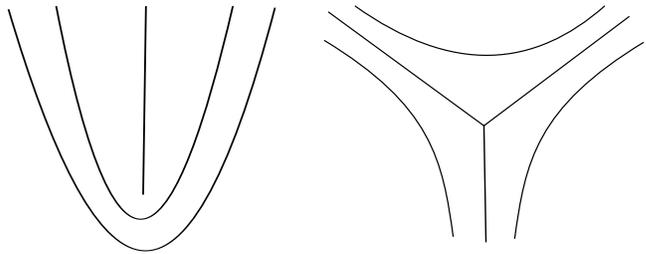}
\caption{Illustrations of the two types of half-index singularities for ridge 
systems on a two-dimensional space. 
On the left is a loop singularity with index $i = 1/2$, and on the
right is a triradius with $i = -1/2$.}
\label{fig:HalfSingularities}
\end{figure}

\begin{figure}[tb]
\centering
\includegraphics[width = 3.325 in, keepaspectratio]{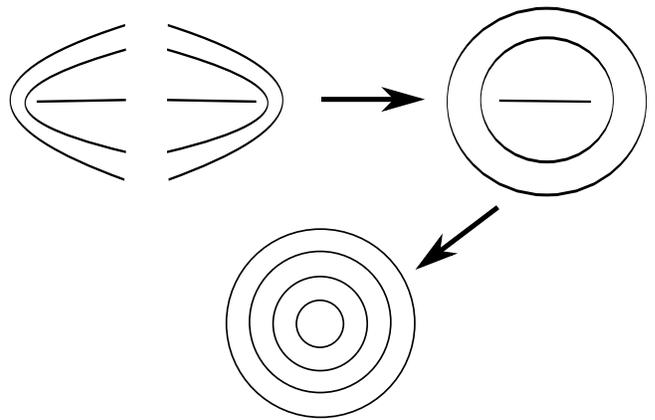}
\caption{An illustration of the formation of a singularity with index $i = 1$ from two 
loop singularities with index $i = 1/2$. 
The local structure of the two loops is shown in the top left, and the arrow
represents, schematically, how they might join together into the extended 
pattern at the top right. 
Finally, the two loop singularities can be brought together until they coincide
(which we represent by an arrow pointing to the image at the bottom). 
This resulting local pattern can be assigned an orientation and is equivalent 
to the singular point of a vector field.}
\label{fig:WedgetoCircle}
\end{figure}

\begin{figure}[tb]
\centering
\includegraphics[width = 3.325 in, keepaspectratio]{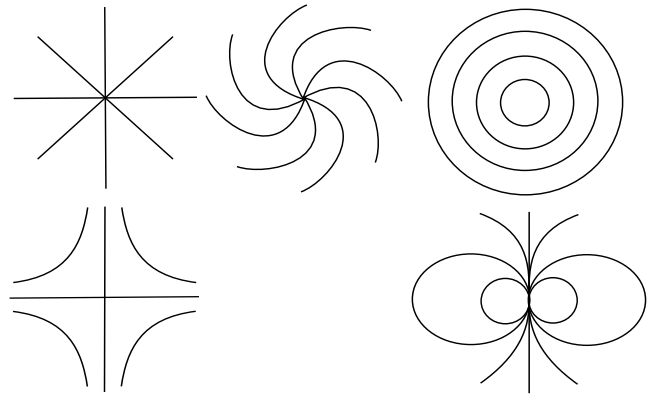}
\caption{Top row: Diagrams of three orientable ridge patterns, which can be 
made from a combination of two loops, all with index $i = 1$.
Bottom row: On the left is an orientable ridge pattern with index $i = -1$
(which is identical to a saddle point of a vector field). 
It can be constructed by joining two triradii singularities.
The figure on the right shows a dipolelike pattern with index $i=2$, which can 
come from the coincidence of four loops.}
\label{fig:AllIndex}
\end{figure}

The concept of an index and the formal statement of the Poincar\'e-Hopf
theorem generalizes naturally to ridge systems, fields of unoriented lines such as the tendex lines on $S$. 
For ridge systems on the sphere, the index of a singular point can be a half-integer \cite{DelmarcelleHesselink1994}. 
Intuitively, this can occur because fields of lines do not have orientation.
As one traverses counterclockwise about a small circle around a singular 
point, the local pattern of lines can rotate 
through an angle of $\pm \pi$ during the transit. 
We illustrate the two fundamental types of singularity in Fig.\
\ref{fig:HalfSingularities}, which, following \cite{PenroseRindler2}, we call 
loops for index $ i = 1/2$ and triradii for $i = -1/2$. 
One can argue that the Poincar\'{e}-Hopf Theorem holds for ridge systems, by 
noting that we can create a singular point with integer index by bringing two half-index singularities together 
(see Fig.\ \ref{fig:WedgetoCircle} for a schematic of the creation of a singularity of
index $i=1$ from two loop singularities).
Ridge patterns near singularities with integer index $i=\pm 1$ can be 
assigned orientations consistently; they must, therefore, have the same 
topological properties as streamlines of vector fields (which, in turn, have 
the same properties as the underlying vector fields themselves).
By arguing that one can always deform a ridge system so that its singular 
points have integer index, one can see that the sum of the indices of a ridge 
system on a sphere must equal the Euler characteristic of the surface, 
$\chi = 2$
(see \cite{DelmarcelleHesselink1994} and the references therein for a more 
formal statement and proof of this theorem). 
In Fig.\ \ref{fig:AllIndex} we show several other ridge singularities with 
integer index for completeness.
In the top row, we show three patterns with index $i=1$, and in the bottom 
left, we sketch a saddle type singularity with index $i=-1$. 
All of these patterns can be consistently assigned an orientation
and, thus, have the same topological properties as vector field singularities.

Having arrived at the result that the tendex lines on $S$ must have singular 
points in a general, asymptotically flat vacuum spacetime, we now recall the 
fact that the singular points appear where there is a degenerate eigenvalue
of the tidal tensor. From the result of Sec. \ref{sec:GravWaves}, the only degeneracies occur where the curvature vanishes completely, and it follows therefore that there must be points of vanishing curvature on $S$. 
In general we would expect the radiation to vanish at a minimum of four 
points, as Penrose \cite{Penrose1965} had previously noted. In this case there would be four loop singularities with index $i = 1/2$, whose index sums to $\chi =2$.
As we highlight in Sec.\ \ref{sec:Examples}, where we show several examples of 
multipolar radiation in linearized theory, the number of singular points,
the types of singularities, and the pattern of the tendex lines contains additional information.

Additional symmetry, however, can modify the structure of the singular points,
as we see in the simple example of an axisymmetric, head-on collision of two 
nonspinning black holes.
Axisymmetry guarantees that the Weyl scalar $\Psi_4$ is purely real when we 
construct our tetrad, Eq.\ \eqref{OrthonormalBasis}, by choosing ${\bf e}_2$ 
and ${\bf e}_3$ to be the orthonormal basis vectors of spherical 
polar coordinates on $S$, $\bm{e}_\theta$ and $\bm{e}_\phi$ \cite{Fiske2005}. 
Using the relation $\Psi_4 = -\ddot{h}_+ + i \ddot{h}_\times$, we see that the 
waves are purely $+$ polarized.
By substituting this relationship into \eqref{PlaneWaveQ}, we also see that 
$\bm{e}_\theta$ and $\bm{e}_\phi$ are the eigenvectors whose integral curves 
are the tendex lines. 
The tendex lines, therefore, are the lines of constant latitude and longitude, 
and the singular points reside on the north and south poles of $S$.
Their index must be $i=1$, and the local pattern at the singularity will 
resemble the pattern at the top left of Fig.\ \ref{fig:AllIndex} for one 
set of lines, and the image on the top right of Fig.\ \ref{fig:AllIndex} 
for the other set (see also \cite{PRD1}).
In this special situation, axisymmetry demands that there be two singular 
points on the axis, rather than four (or more).
Moreover, these singular points are each generated from the coincidence of two 
loop singularities, with one singular point at each end of the axis of symmetry. 
Similarly, if $\Psi_4$ were purely imaginary, then the radiation would only 
contain the $\times$ polarization.
The $\pi/4$ rotations of the unit spherical vectors would then be the eigenvectors of 
the tidal field, and the two singularities at the poles would resemble that 
illustrated at the top middle of Fig.\ \ref{fig:AllIndex}.

It is even conceivable that four loops could merge into one singular point.
This singularity would have the dipolelike pattern illustrated at the bottom 
right of Fig.\ \ref{fig:AllIndex}, and it would have index $i=2$. 
Though this situation seems very special, we show in the next section how a 
finely tuned linear combination of mass and current multipoles can give rise 
to this pattern.
Because there is only one zero, the radiation is beamed in the direction
opposite the zero, resulting in a net flux of momentum opposite
the lone singular point.


Before concluding this section, we address two possible concerns. 
The scalar $\Psi_4=C_{\mu\nu\rho\sigma} n^\mu \bar m^\nu n^\rho \bar m^\sigma$ 
depends both on the curvature and on the chosen tetrad. 
We first emphasize that the singular points we have discussed have nothing to 
do with tetrad considerations, in particular with the behavior of the vectors 
tangent to the sphere, $\bm m$, and $\bm{\bar m}$. 
Though these vectors will also become singular at points on the sphere, we are 
free to use a different tetrad on $S$ in these regions, just as we can 
cover the sphere everywhere with smooth coordinates using overlapping charts. 
Secondly, the vanishing of radiation does not occur due to the null vector 
$\bm n$ coinciding with a principle null direction of the spacetime. 
We note that, if $\Psi_4$ vanishes at a point on $S$, then a change of basis 
cannot make $\Psi_4$ (or any of the other curvature scalars) nonvanishing. 
For example, a rotation about $\bm l$ by a complex parameter $a$ induces a 
transformation on the other basis vectors,
\begin{eqnarray}
{\bm l}' & =& {\bm l} \,, \nonumber \\
{\bm m}' & =&  {\bm m} + a {\bm l} \,, \nonumber \\
{\bm{ \bar m}}' & = &  {\bm{ \bar m}} + \bar a {\bm l} \, ,\nonumber \\
{\bm n}' & = & {\bm n}+ a {\bm{ \bar m}} + \bar a {\bm m} + a \bar a {\bm l} \,.
\end{eqnarray}
Under this rotation, $\Psi_4$ transforms as
\begin{eqnarray}
\Psi_4' & = & \Psi_4 + 4 \bar a \Psi_3 + 6 \bar a^2 \Psi_2 + 4 \bar a^3 \Psi_1 + \bar a^4 \Psi_0 \,,
\end{eqnarray}
which vanishes when the Weyl scalars are zero in the original basis. 
The remaining scalars transform analogously, and the other independent tetrad 
transformations are also homogeneous in the Weyl scalars 
(see e.g.\ \cite{Stephani}).

\section{Examples from Linearized Gravity}
\label{sec:Examples}

We now give several examples of the tendex and vortex patterns on $S$ from 
weak-field, multipolar sources. 
We first investigate quadrupolar radiation, produced by a time-varying 
quadrupole moment. 
For many astrophysical sources, such as the inspiral of comparable mass 
compact objects, the gravitational radiation is predominantly quadrupolar.
As a result, our calculations will capture features of the radiation 
coming from these astrophysical systems. 
We will then study a combination of rotating mass- and current-quadrupole moments that are 
phase-locked.
The locking of these moments was observed by Schnittman et al.\ in their
multipolar analysis of the extreme-kick merger \cite{Schnittman2008}.
We conclude this section by discussing isolated higher multipoles.
Although it is unlikely that astrophysical sources will contain only 
higher multipoles, it is of interest to see what kinds of tendex patterns
occur.
More importantly, while the tidal tensor is a linear combination of multipoles,
the tendex lines will depend nonlinearly on the different moments.
Actual astrophysical sources will contain a superposition of multipoles, and
it is important to see how superpositions of multipoles change the 
leading-order quadrupole pattern.

We perform our calculations in linearized theory about flat space, and we 
use spherical polar coordinates and their corresponding unit vectors for our 
basis. 
One can compute from the multipolar metric in \cite{Thorne1980} that for a 
symmetric, trace-free (STF) quadrupole moment $\mathcal{I}_{ab}$, the leading-order contributions to $\bm{\mathcal E}$ and $\bm{\mathcal B}$ on $S$ are
\ba
\label{eq:mass_tidal}
\mathcal E_{ab}^{(\ell=2)} & = & -\frac{1}{2r} \left[ {}^{(4)} 
\mathcal I_{ab}(t-r) + \epsilon_{ac} {}^{(4)}\mathcal I_{cd}(t-r)
\epsilon_{db}\right]^{\rm TT}, \nonumber \\
\\
\label{eq:mass_drag}
\mathcal B_{ab}^{(\ell=2)} & = & -\frac 1r \left[\epsilon_{c(a} {}^{(4)}
\mathcal I_{b)c}(t-r)\right]^{\rm TT} \, .
\ea
Here, the superscript $^{(4)}$ indicates four time derivatives, TT means to 
take the transverse-traceless projection of the expression, and $\epsilon_{ac}$
is the antisymmetric tensor on a sphere. 
In this expression, and in what follows, the Latin indices run only over the basis 
vectors $\bm{e}_\theta$ and $\bm{e}_\phi$, and repeated Latin indices are summed over even when they are both lowered.

\subsection{Rotating Mass Quadrupole}
\label{MassQuad}

As our first example, we calculate the STF quadrupole
moment of two equal point masses (with mass $M/2$) separated by a distance $a$
in the equatorial plane, and rotating at an orbital frequency $\Omega$.
We find that
\begin{eqnarray}
\nonumber
{}^{(4)}\mathcal I_{\theta\theta}(t-r)  
& = &  Ma^2\Omega^4(1+\cos^2\theta)\cos\{2[\phi-\Omega(t-r)]\}\, ,\\
\nonumber
{}^{(4)}\mathcal I_{\theta\phi}(t-r)  
& = & -2Ma^2\Omega^4\cos\theta\sin\{2[\phi-\Omega(t-r)]\} \, ,\\
{}^{(4)}\mathcal I_{\phi\phi}(t-r) & = & 
-{}^{(4)}\mathcal I_{\theta\theta}(t-r) \, .
\label{eq:mass_quadrupole}
\end{eqnarray}
By substituting these expressions into Eqs.\ \eqref{eq:mass_tidal} and 
\eqref{Eigenvecs}, we find the eigenvectors of the tidal field. 
We can then calculate the tendex lines on the sphere by solving 
Eq.\ \eqref{StreamlineEqn} with a convenient normalization of the parameter along the curves,
\begin{eqnarray}
\label{eq:tendex_lines_th}
\frac{d\theta}{ds} &=& \frac 1r {}^{(4)}\mathcal I_{\theta\phi} \, ,\\
\frac{d\phi}{ds} &=& \frac 1{r\sin\theta}(- {}^{(4)}\mathcal I_{\theta\theta}
-\lambda_+) \, .
\label{eq:tendex_lines_ph}
\end{eqnarray} 
Here, $\lambda_+$ is the positive eigenvalue. 
The differential equation for the vortex lines [found from the corresponding
frame-drag field of Eq.\ \eqref{eq:mass_drag}], has the same form as those of 
the tendex lines above; however, one must replace 
${}^{(4)}\mathcal I_{\theta\phi}$ in the first equation by 
${}^{(4)}\mathcal I_{\theta\theta}$ and ${}^{(4)}\mathcal I_{\theta\theta}$
by $-{}^{(4)}\mathcal I_{\theta\phi}$ in the second equation.

\begin{figure}[t]
\includegraphics[width=0.9\columnwidth]{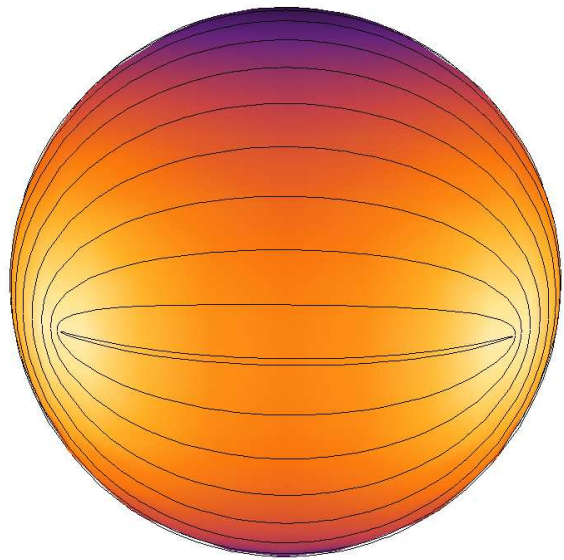}
\caption{(color online). 
The positive tendex lines on $S$ generated by a rotating quadrupole moment
in linearized gravity. 
The sphere is colored by the magnitude of the eigenvalue; purple 
(darker) areas at the poles corresponding to a large eigenvalue, and yellow 
(lighter) areas near the equator indicate a value closer to zero.
Four loop singularities appear equally spaced on the equator at the 
points of vanishing tendicity.}
\label{fig:MassQuadTendex}
\end{figure}

\begin{figure}[t]
\includegraphics[width=0.9\columnwidth]{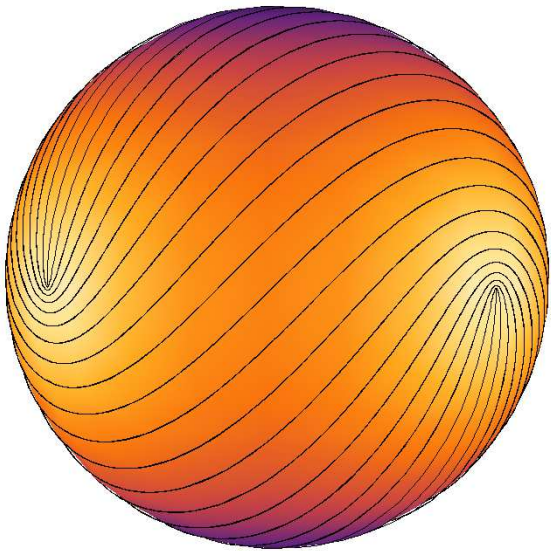}
\caption{(color online).
As in Fig. \ref{fig:MassQuadTendex}, we show the positive vortex lines and 
their magnitude of the eigenvalue on $S$ (using the same coloring as in
that figure). 
The loop singularities lie at the same locations as they do for the
tendex lines, but they are locally rotated by $\pi/4$.}
\label{fig:MassQuadVortex}
\end{figure}

We show the tendex and vortex lines corresponding to the positive 
eigenvalues in Figs.\ \ref{fig:MassQuadTendex} and \ref{fig:MassQuadVortex}, 
respectively, at a retarded time $t-r=0$. 
We also plot the magnitude of the eigenvalue on the sphere, using a color
scheme in which purple (darker) regions at the poles correspond to large 
eigenvalues and yellow (lighter) colors near the equator are closer to zero.
Both the tendex and vortex lines have four equally spaced loop singularities 
on the equator at the points where the field is zero (the two on the back
side of the sphere are not shown). 
Because the vortex and tendex lines must cross each other at an angle of 
$\pi/4$, the global geometric patterns are quite different.

We note here that these two figures also provide a visualization for the
transverse-traceless, ``pure-spin'' tensor spherical harmonics 
\cite{Thorne1980}. 
For example, we can see that the mass-quadrupole tendex lines are the integral 
curves of the eigenvectors of the real part of the $\ell = 2, \  m = 2$ 
electric-type transverse-traceless tensor harmonic. 
First, the tendex lines correspond to the electric-type harmonic, because 
the tidal tensor is even under parity.
Second, the radiation pattern will not contain an $\ell = 2, \ m=0$ harmonic,
because the overall magnitude of the quadrupole moment of the source is not 
changing in time; also, the $\ell =2, \ m=\pm 1$ harmonics are absent because 
the source is an equal-mass binary and is symmetric under a rotation of $\pi$. 
Finally, the $\ell =2, \ m=-2$ moment is equal in magnitude to the $m=2$ 
harmonic, since the tidal tensor is real. 
By similar considerations, we can identify the vortex lines of the mass 
quadrupole as a visualization of the real parts of the $\ell = 2, \ m=2 $ 
magnetic-type tensor harmonics.

In addition, the eigenvalue (the identical color patterns of 
Figs.\ \ref{fig:MassQuadTendex} and \ref{fig:MassQuadVortex}) is given by the 
magnitude of the sum of spin-weighted spherical harmonics, 
\be
\lambda_+\propto|{}_{-2}Y_{22} + {}_{-2}Y_{2-2}| \,.
\ee
One can see this most easily by using the symmetries described above, the
expression for the eigenvalue $\lambda_+=|\Psi_4|/2$, and the spin-weighted
spherical harmonic decomposition of $\Psi_4$.
It is also possible to verify this expression using the tensor harmonics
above and the standard relations between tensor spherical harmonics and 
spin-weighted spherical harmonics (see e.g.\ \cite{Thorne1980}). 
Radiation from numerical spacetimes is usually decomposed into spin-weighted 
spherical harmonics, and, as a result, the pattern of the eigenvalue is 
familiar.
The tendex lines, however, also show the polarization pattern of the waves on 
$S$ (a feature that numerical simulations rarely explicitly highlight). 
Figure\ \ref{fig:MassQuadTendex} (and the accompanying negative-tendicity lines 
not shown) gives the directions of preferred strain on $S$, and hence the wave 
polarization that can be inferred from gravitational-wave-interferometer 
networks such as LIGO/VIRGO. 
Thus, visualizations such as Fig.\ \ref{fig:MassQuadTendex} give complete
information about the gravitational waves passing through $S$.

\subsection{Rotating Mass and Current Quadrupoles in Phase}
\label{SupremeKick}

As our second example, we will consider a source that also has a time-varying 
current-quadrupole moment, $\mathcal S_{ab}$.
In linearized theory, one can show that the tidal tensor and frame-drag
field of a current quadrupole are simply related to those of a mass
quadrupole.
In fact, $\mathcal B_{ab}$ of the current quadrupole has exactly the
same form as $\mathcal E_{ab}$ of a mass quadrupole, Eq.\ 
(\ref{eq:mass_tidal}), when one replaces ${}^{(4)}\mathcal I_{ab}$ by 
$(4/3) {}^{(4)}\mathcal S_{ab}$.
Similarly, $\mathcal E_{ab}$ of the current quadrupole is identical to
$\mathcal B_{ab}$ of a mass quadrupole, Eq.\ (\ref{eq:mass_drag}), when
${}^{(4)}\mathcal I_{ab}$ is replaced by $-(4/3) {}^{(4)}\mathcal S_{ab}$.

We impose that the source's mass- and current-quadrupole moments rotate 
in phase, with frequency $\Omega$, and with the current quadrupole lagging in 
phase by $\pi/2$.
This arrangement of multipoles models the lowest multipoles during the merger 
and ringdown of the extreme-kick configuration (a collision of equal-mass 
black holes in a quasicircular orbit that have spins of equal magnitude lying in the orbital plane, but pointing in opposite directions), when the
mass- and current-quadrupole moments rotate in phase \cite{Schnittman2008}.
The relative amplitude of the mass- and current-multipoles depends upon,
among other variables, the amplitude of the black-holes' spin.
We, therefore, include a free parameter $C$ in the strength of the current 
quadrupole which represents the effect of changing the spin. 
An order-of-magnitude estimate based on two fast-spinning holes orbiting near 
the end of their inspiral indicates that that their amplitudes could be 
nearly equal, $C = O(1)$. 
To determine the exact relative amplitude of the mass- and current-quadrupole 
moments of the radiation would require comparison with numerical relativity
results. 

We calculate the current-quadrupole moment by scaling the
mass quadrupole by the appropriate factor of $C$ and letting the term 
$2[\phi-\Omega(t-r)]$ in the equations for $\mathcal I_{ab}(t-r)$ become 
$2[\phi-\Omega(t-r)]-\pi/2$ in the corresponding expressions for 
$\mathcal S_{ab}(t-r)$.
In linearized theory, the tidal tensor and frame-drag fields of the different
multipoles add directly.
As a result, the equations for the tendex lines have the same form as Eqs.\
\eqref{eq:tendex_lines_th} and \eqref{eq:tendex_lines_ph}, but one must now 
replace the mass quadrupole ${}^{(4)}\mathcal I_{\theta\phi}$ by 
${}^{(4)}\mathcal I_{\theta\phi} - (4/3) {}^{(4)}\mathcal S_{\theta\theta}$
in the first expression and ${}^{(4)}\mathcal I_{\theta\theta}$ by
${}^{(4)}\mathcal I_{\theta\theta} + (4/3) {}^{(4)}\mathcal S_{\theta\phi}$
in the second.

\begin{figure}[t]
\includegraphics[width=0.9\columnwidth]{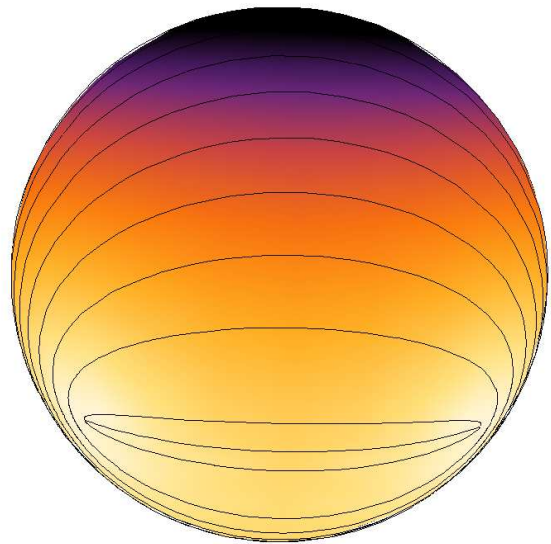}
\caption{(color online).
The positive tendex lines on $S$ generated by the superposition of rotating 
mass- and current-quadrupole moments, $\pi/2$ out of phase, in linearized 
gravity.
The sphere is colored by the tendicity as in Fig.\ \ref{fig:MassQuadTendex}. 
Interference between the moments leads to beaming of the radiation toward
the north pole.
Similarly, the singular points of the tendex lines now fall on a line of 
constant latitude in the southern hemisphere.}
\label{fig:kick}
\end{figure}

First, we allow the current quadrupole to be half as large as the mass
quadrupole, $C=1/2$. We show the positive tendex lines and positive eigenvalue in Fig.\ \ref{fig:kick}. Because of the relative phase and amplitude of the two moments, the tensors 
add constructively in the northern hemisphere and destructively in the 
southern hemisphere on $S$.
This is evident in the eigenvalue on the sphere in Fig.\ \ref{fig:kick},
which, one can argue, is now given by an unequal superposition of
spin-weighted spherical harmonics, 
\be
\lambda_+ \propto |{}_{-2}Y_{22} + b{}_{-2}Y_{2-2}| \,,
\ee
with $b < 1$.
As in previous figures, dark colors (black and purple) represent where
the eigenvalue is large, and light colors (white and yellow) show
where it is nearly zero.
While the singular points are still equally spaced on a line of constant
latitude, they no longer reside on the equator; they now fall in the
southern hemisphere.
This is a direct consequence of the beaming of radiation toward the
northern pole.

\begin{figure}[t]
\includegraphics[width=0.9\columnwidth]{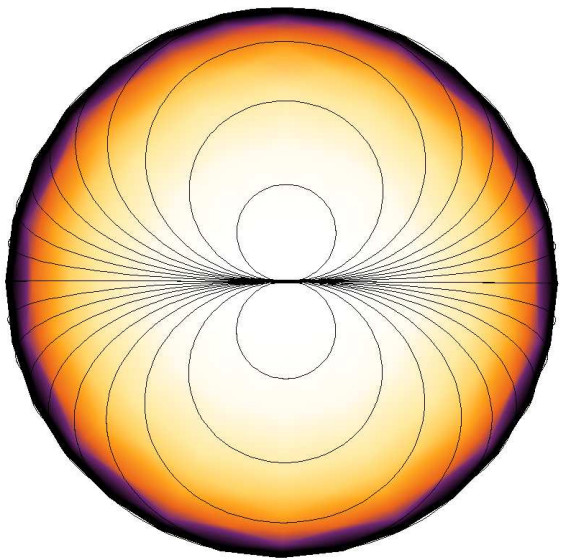}
\caption{(color online). South polar region of the tendex line pattern of a 
gravitational wave generated by rotating mass- and current-quadrupole moments. 
The amplitude and phase of the moments are chosen so that the radiation 
vanishes only at the south pole. 
The purple (darker) areas indicate a positive eigenvalue, while yellow 
(lighter) areas are values closer to zero. 
The singularity at the pole has index $i=2$.}
\label{fig:kick2}
\end{figure}

The case shown above has strong beaming, but it is possible to make
the beaming more pronounced. To get the greatest interference of the multipoles, the mass and current
quadrupoles must have equal amplitude in the tidal field.
Because the tidal field of the current quadrupole is $4/3$ as large as 
the tidal field of the mass quadrupole, setting $C=3/4$ gives the strongest constructive interference in the tidal
fields.
In this case, the eigenvalue vanishes at just one point, the south pole, and the eigenvalue can be shown to be proportional to just a single
spin-weighted spherical harmonic,
\be
\lambda_+ \propto|{}_{-2}Y_{22}| \,.
\ee
As a result, the four equally spaced singular points of the tendex lines
must coincide at one singular point whose index must be $i=2$.
This is precisely the dipolelike pattern depicted in Fig.\ 
\ref{fig:AllIndex}.
We show the tendex lines around the south pole in Fig.\ \ref{fig:kick2}.
The vortex lines are identical to the tendex lines, but they are globally
rotated by $\pi/4$ in this specific case.

We see that the beaming can be maximized by carefully tuning the phase and 
amplitude of the mass- and current-quadrupole moments. 
Interestingly, the maximally beamed configuration corresponds with 
the coincidence of all singular points at the south pole in the radiation zone. 
Whether this degree of beaming could occur from astrophysical sources is 
an open question.

\subsection{Higher Multipoles of Rotating Point Masses}

We also investigate the effect of including higher multipoles on the tendex
lines on $S$.
For the orbiting, nonspinning, point masses of the first example
the next two lowest multipoles arise from the current octopole
(the $\ell=3$ STF moment \cite{Thorne1980})
and the mass hexadecapole (the $\ell=4$ STF moment).
From the multipolar metric in \cite{Thorne1980}, one can show that the tidal
field for these two moments are
\begin{eqnarray}
\mathcal E_{ab}^{\ell=3} & = & -\frac 1{2r} \left[\epsilon_{c(a} {}^{(5)}
\mathcal S_{b)c r}(t-r) \right]^{\rm TT} \, ,\\
\mathcal E_{ab}^{\ell=4} & = & -\frac{1}{24 r} \left[ {}^{(6)}
\mathcal I_{ab r r}(t-r) + \epsilon_{ac} \epsilon_{db} {}^{(4)}
\mathcal I_{cd r r}(t-r) \right]^{\rm TT} \, ,
\nonumber \\
\end{eqnarray}
where the index $r$ indicates contraction with the radial basis vector 
${\bm e}_r$, and so repeated $r$ indices do not indicate summation. 
The STF current-octopole moment, can be expressed compactly
as $\mathcal S_{ijk} = (L_N^i x_A^j x_A^k)^{\rm STF}$, where $L_N^i$ is the
Newtonian angular momentum and $x_A^j$ is the position of one of the point
masses. The superscript STF indicates that all indices should be symmetrized, and all traces removed. In Cartesian coordinates, the vectors have the simple forms
${\bf L}_N = (0,0,Mav/4)$ and
${\bf x}_A = (a/2)(\cos[\Omega t],\sin[\Omega t],0)$,
where $\Omega$ is the Keplerian frequency and $v$ is the relative velocity.
Similarly, one can write the STF mass-hexadecapole moment
as $\mathcal I_{ijkl} = M(x_A^i x_A^j x_A^k x_A^l)^{\rm STF}$, for the same
vector $x_A^j$ as above.
Because these tensors have many components, we shall only list those that
are relevant for finding the tendex lines.
We will also define $\alpha=\phi-\Omega(t-r)$ for convenience.
For the current octopole the relevant components are
\begin{eqnarray}
\nonumber
{}^{(5)}\mathcal S_{\theta\theta r}(t-r)
& = &  -\frac{Ma^3v\Omega^5}{24}(5\cos\theta+3\cos 3\theta)
\sin 2\alpha \, ,\\
\nonumber
{}^{(5)}\mathcal S_{\theta\phi r}(t-r)
& = & -\frac{Ma^3v\Omega^5}3\cos 2\theta\cos 2\alpha \, ,\\
{}^{(5)}\mathcal S_{\phi\phi r}(t-r) & = &
-{}^{(5)}\mathcal S_{\theta\theta r}(t-r) \, ,
\label{eq:current_octopole}
\end{eqnarray}
and for the mass hexadecapole they are
\begin{eqnarray}
\nonumber
{}^{(6)}\mathcal I_{\theta\theta  r r}(t-r)
& = &  \frac{Ma^4\Omega^6}8[(\cos^2\theta+\cos 4\theta)
\cos 2\alpha \\
\nonumber
&& -128\sin^2\theta(1+\cos^2\theta) \cos 4\alpha]\, ,\\
\nonumber
{}^{(6)}\mathcal I_{\theta\phi r r}(t-r)
& = & -\frac{Ma^4\Omega^6}4 [\cos 3\theta \sin 2\alpha \\
\nonumber
&& - 128 \sin^2\theta\cos\theta \sin 4\alpha]\, ,\\
{}^{(6)}\mathcal I_{\phi\phi r r}(t-r) & = &
-{}^{(6)}\mathcal I_{\theta\theta r r}(t-r) \, .
\label{eq:mass_hexadecapole}
\end{eqnarray}
The tendex lines of the current octopole can be found by solving the system
of differential equations in Eqs.\ \eqref{eq:tendex_lines_th} and 
\eqref{eq:tendex_lines_ph} by substituting
${}^{(4)}\mathcal I_{\theta\phi}$ by
${}^{(5)}\mathcal S_{\theta\theta r}/2$ and
${}^{(4)}\mathcal I_{\theta\theta}$ by
$-{}^{(5)}\mathcal S_{\theta\phi r}/2$
Similarly, for the mass hexadecapole, one must make the substitutions of
${}^{(4)}\mathcal I_{\theta\phi}$ by
${}^{(6)}\mathcal I_{\theta\phi r r}/12$ and
${}^{(4)}\mathcal I_{\theta\theta}$ by
${}^{(6)}\mathcal I_{\theta\theta r r}/12$ in the same equations.

\begin{figure}
\includegraphics[width=0.9\columnwidth]{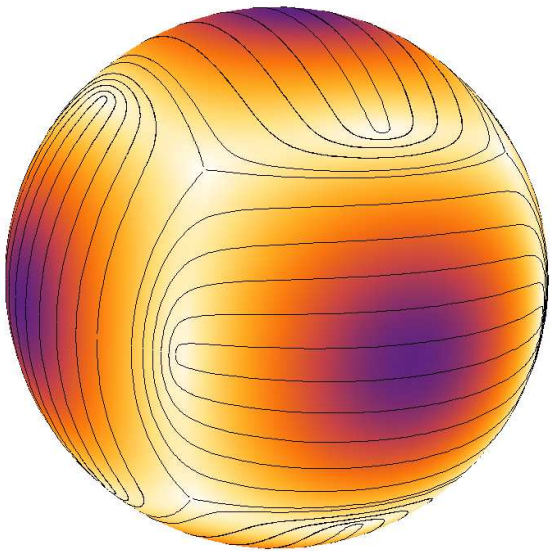}
\caption{(color online). 
The tendex lines of a current-octopole moment of an
equal-mass, circular binary of point masses.
The colors on the sphere represent the tendicity, with the same scale
described in Fig.\ \ref{fig:MassQuadTendex}.
The current octopole also has four loop singularities on the equator (at 
the same position of those of the rotating quadrupole), but it has eight 
additional loops and eight triradius singularities off of the equator.
Only half of the singular points are visible on the sphere; the other
half appear on the back side.}
\label{fig:octopole}
\end{figure}

\begin{figure}
\includegraphics[width=0.9\columnwidth]{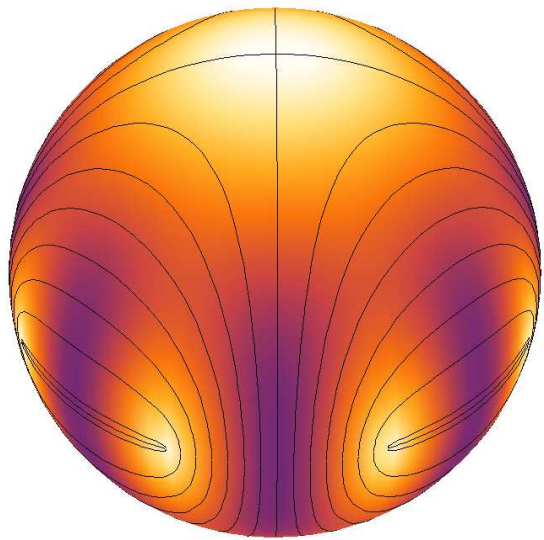}
\caption{(color online). 
The tendex lines on $S$ of the mass hexadecapole of an equal-mass, circular 
binary of point masses, with the sphere colored by the tendicity as in Fig.\ 
\ref{fig:MassQuadTendex}.
The hexadecapole has eight loop singularities equally spaced on the equator
and two saddle-point-like singularities (from the coincidence of two triradius
singularities at a point) at the poles.
Again, only half are visible in the figure.
Four of the singular points on the equator coincide with those of the 
quadrupole, but the remaining four appear at different locations.}
\label{fig:hexadecapole}
\end{figure}

In Fig.\ \ref{fig:octopole} we show the tendex line pattern for the current 
octopole, and in Fig.\ \ref{fig:hexadecapole} we show the pattern for the 
mass hexadecapole. 
Together with the mass quadrupole, Fig.\ \ref{fig:MassQuadTendex}, these are 
the three lowest multipole moments for the equal-mass circular binary. 
For the current octopole, there are eight triradius singular points and 12 loop singularities (and thus the net index is two).
Four of the loop singularities remain equally spaced on the equator,
at the same position of those of the quadrupole, but the remaining 
singularities appear at different points on $S$.
The mass hexadecapole has eight loop singularities equally
spaced on the equator, and there are integer-index saddle-point-like singularities at each pole.

Gravitational radiation from astrophysical sources will likely not be
dominated by these higher multipoles.
Nevertheless, these figures are of interest as examples of tendex lines with 
many singular points and as visualizations of tensor harmonics. 
By analyzing the symmetries in a way analogous to that discussed in 
Sec.\ \ref{MassQuad}, we can identify the current-octopole tendex lines with 
the integral curves of $\ell = 3, \ m=2$ magnetic-type harmonics, and we can
associate the mass-hexadecapole lines with those of the $\ell = 4,\  m=4$ 
electric-type harmonics. 
In the case of the mass hexadecapole, the $m = 2 $ moment is not ruled out by 
symmetry, but it is suppressed relative to the $m = 4$ moment. 
This occurs because the $m=4$ moment oscillates at twice the frequency of 
the $m=2$ moment, and the tidal tensor for this higher-order moment is given 
by taking six time derivatives of the STF moment, 
Eq.\ \eqref{eq:mass_hexadecapole}. 
This enhances the $m=4$ radiation by a factor of $2^6$ over the $m=2$ 
contribution. 
Similarly, we can relate the eigenvalue to the magnitude of the 
corresponding sum of $s=-2$ spin-weighted spherical harmonics, and the tendex 
line patterns to the the polarization directions that could be inferred from
networks of gravitational-wave interferometers.

\begin{figure}
\includegraphics[width=0.9\columnwidth]{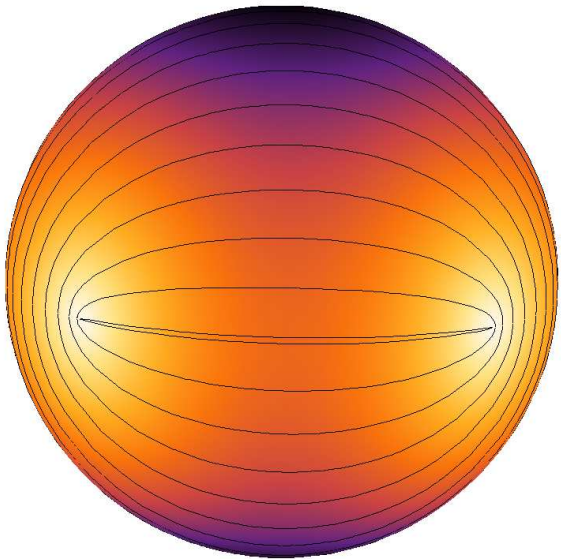}
\caption{(color online). 
The tendex lines of a superposition of mass-quadrupole, current-octopole,
and mass-hexadecapole moments of an equal-mass circular binary.
It assumes a total mass $M=1$, a separation $a=15$, an orbital frequency
$\Omega=a^{-3/2}$ and a velocity $v=a\Omega=a^{-1/2}$.
The sphere is colored by the tendicity in an identical way to that
of  Fig.\ \ref{fig:MassQuadTendex}.
When the tidal tensors of the three multipoles are combined, the net pattern
is dominated by the quadrupole and contains only the four loops.
The loop singularities are no longer equally spaced on the
equator; the two pairs are pushed closer together due to the influence
of the higher multipoles.}
\label{fig:superposition}
\end{figure}

Finally, we show the pattern generated from the linear combination of the 
three lowest multipole moments in Fig.\ \ref{fig:superposition}.
Any astrophysical source will contain several multipoles, with the 
quadrupole being the largest.
The tendex lines depend nonlinearly on the multipoles, and it is important,
therefore, to see to what extent higher multipoles change the overall pattern.
We find the total tidal tensor by linearly combining the tidal tensor of each 
individual moment, and we then find the eigenvectors and tendex lines of the 
total tidal tensor. 
The pattern formed from the combination of multipoles depends upon the
parameters of the binary; in making this figure we assumed (in units in
which $M=1$) a separation of $a=15$, an orbital frequency $\Omega=a^{-3/2}$,
and a velocity $v=a\Omega=a^{-1/2}$. 
When these higher moments are combined with the mass quadrupole, the
tendex line structure resembles that of the mass quadrupole.
The pattern is deformed slightly, however, by the presence of the higher
multipoles. 
The loop singularities on the equator are no longer evenly spaced; rather, 
the pair illustrated (and the corresponding pair which is not visible) are 
pushed slightly closer together.

\section{Conclusions}
\label{sec:Conclusions}

Tendex and vortex lines provide a new tool with which to visualize and study 
the curvature of spacetime. Fundamentally, they allow for the visualization of the Riemann tensor, through its decomposition into two simpler, trace-free and symmetric spatial tensors. These tensors, $\bm{\mathcal E}$ and $\bm{\mathcal B}$, can be completely characterized by their eigenvectors and corresponding eigenvalues. The integral curves of these eigenvector fields are easily visualized, and their meaning is well understood; physically, the lines can be interpreted in terms of local tidal 
strains and differential frame-dragging. Here, the simple nature of these lines allows us to apply well-known topological theorems to the study of radiation passing through a sphere near null infinity.

Tendex line patterns must develop singularities (and thus have vanishing 
tendicity) on a closed surface.
When we applied this fact to the tendex lines of gravitational radiation 
near null infinity from arbitrary physical systems, we could easily show
that the gravitational radiation must at least vanish in isolated directions.
Although this result is somewhat obvious in retrospect and has been noted 
before \cite{Penrose1965}, the result does not appear to be well-known. 
We also began exploring the manner in which these singular points can provide 
a sort of fingerprint for radiating spacetimes. 
The essential elements of this fingerprint consist of the zeros of the 
curvature on the sphere, together with the index and the tendex line pattern around these 
zeros. 
We studied these patterns for a few specific examples, such as the four 
equally spaced loops of a rotating mass quadrupole.
A more interesting case is that of a radiating spacetime composed of locked, 
rotating mass and current quadrupoles, which can be thought of as a simplified
model of the late stages of the extreme-kick black-hole-binary merger.
Here, the shifted positions of the singular points of the tendex pattern 
provide a direct illustration of gravitational beaming for this system. 
By seeking the most extreme topological arrangement of singular points, we 
also described a maximally beaming configuration of this system. 

The radiation generated by higher-order STF multipole moments gives more 
complex examples of tendex and vortex patterns, with many singular points of 
varied types. 
Additionally, we argued that their tendex and vortex patterns provide a 
visualization of the tensor spherical harmonics on the sphere; the eigenvalue 
illustrates the magnitude of these harmonics, and the lines show the tensor's 
polarization in an intuitive manner. 
The sum of the three multipoles illustrated in Fig.\ \ref{fig:superposition} 
shows how including higher-order multipoles slightly deforms the pattern of
quadrupole radiation to make a more accurate total radiation pattern of the 
equal-mass binary. 
Similar illustrations of complete radiation patterns could be readily produced 
from numerical spacetimes, when $\Psi_4$ is extracted asymptotically using a 
tetrad with appropriate peeling properties. 
Such visualizations, and their evolution in time, could provide a useful 
method for visualizing the gravitational emission from these systems.

This study of the tendex and vortex lines (and their singular points) of 
asymptotic radiation fields is one of several \cite{PRD1} exploring and
developing this new perspective on spacetime visualization.
Naturally, it would be of interest to extend the two-dimensional case here
to a larger study of the singular points in the full, three-dimensional 
tendex and vortex fields. Methods to find and visualize the singular points (and singular lines) of 3D tensors have been discussed preliminarily in \cite{Weickert2006}, though there is still room for further work. We suspect that singular points will be important in visualizing and studying the properties of numerical spacetimes with these methods. Further, we expect that there is still much to be learned from the study of the vortexes and tendexes of dynamical spacetimes.

\begin{acknowledgments} 
We thank Rob Owen for inspiring our investigation of ridge topology in tendex 
and vortex patterns. 
We would also like to thank Yanbei Chen, Tanja Hinderer, Jeffrey D. Kaplan, Geoffrey Lovelace, Charles W. Misner, Ezra T. Newman, and Kip S. Thorne for valuable discussions. This research was supported by NSF Grants No.\ PHY-0601459, PHY-0653653, 
PHY-1005655, CAREER Grant PHY-0956189, NASA Grant No.\ NNX09AF97G, the Sherman 
Fairchild Foundation, the Brinson Foundation, and the David and Barabara Groce 
Startup Fund.
\end{acknowledgments}

\bibliography{References/MyRefs}

\end{document}